\documentstyle[12pt]{article}                                 

\begin{document}

\newcommand {\bea}{\begin{eqnarray}}
\newcommand {\eea}{\end{eqnarray}}
\newcommand {\be}{\begin{equation}}
\newcommand {\ee}{\end{equation}}

\newcommand{\myprime}{^\prime}
\newcommand{\grad}{\nabla}
\newcommand{\mass}{{\cal M}}
\newcommand{\lapse}{N^{t}}
\newcommand{\shift}{N^{r}}

\title{
\begin{flushright}
\begin{small}
hep-th/9609084 \\
UPR-715-T \\
IASSNS-HEP 96/92 \\
September 1996 \\
\end{small}
\end{flushright}
\vspace{1.cm}
Classical Hair in String Theory II:\\ 
Explicit Calculations}

\author{Finn Larsen
\thanks{Research supported in part by 
Danish National Science Foundation.}\\
\small Department of Physics and Astronomy\\
\small University of Pennsylvania\\
\small Philadelphia, PA 19104-6396 \\
\small e-mail: larsen@cvetic.hep.upenn.edu
\and 
Frank Wilczek
\thanks{Research supported in part by  
DOE grant~DE-FG02-90ER40542.}\\
\small School of Natural Sciences\\
\small Institute for Advanced Study\\
\small Princeton, NJ 08540 \\
\small e-mail: wilczek@sns.ias.edu
}

\maketitle

\newpage
\begin{abstract}
After emphasizing the importance of obtaining a space-time
understanding of black hole entropy, we further elaborate our program
to identify the degrees of freedom of black holes with classical
space-time degrees of freedom.  The Cveti\v{c}-Youm dyonic black holes
are discussed in some detail as an example.  In this example hair
degrees of freedom transforming as an effective string can be
identified explicitly.  We discuss issues concerning charge
quantization, identification of winding, and tension renormalization
that arise in counting the associated degrees of freedom.  The
possibility of other forms of hair in this example, and the prospects
for making contact with D-brane ideas, are briefly considered.
\end{abstract}

\newpage

\section{Introduction: State-Counting, Deterministic Evolution,
and Hair}

The problem of understanding black holes as quantum-mechanical
objects, although presumably academic for the forseeable future,
has attracted great attention recently
among theoretical physicists, due primarily to a fundamental
conceptual tension.

The thermodynamic properties of macroscopic
black holes are widely believed to
be universal; that is, essentially independent
of any details of the microscopic theory in which gravity
is embedded.  This belief
is based on their remarkably simple derivation
from very basic properties of the Einstein-Hilbert
action~\cite{hawking75,hawkingtwo,carlip,hair1}. Indeed, the thermodynamic 
parameters of large black holes follow from general covariance,
the causal structure of black hole backgrounds, 
and basic considerations of field theory at small curvature,
which are plausibly independent of
the matter content and other details of the theory.

On the other hand it is notorious that these thermodynamic
properties lead to considerable
conceptual tension with standard principles
of quantum mechanics.  In particular, the possibility that
black holes formed under widely varying initial conditions will all
eventually evaporate into approximately thermal radiation is not
easily reconciled with the existence of a unitary S-matrix
governing the dynamics.  Indeed the evolution of many, possibly
finely structured, initial states all into a common
final state (or density matrix) -- namely
undistinguished thermal
glop -- could not follow consistently from
fundamentally deterministic, time (or PCT) reversible equations for
a closed system.   This has led Hawking~\cite{hawking76},
and others~\cite{bekenstein}, to suggest that
no such equations can exist, and that a new level of indeterminacy,
beyond that of standard quantum mechanics, is
inevitably associated with the existence of black holes.

The strength of these very broad arguments is open to serious doubt,
however.  The thermal character of the radiation was only established
by approximate, semi-classical calculations for large black holes
far from extremality.  Given too naive a literal interpretation it
is
demonstrably false: in particular,
the whole concept of thermal behavior becomes
ambiguous for near-extremal holes~\cite{nperson,susy} and some
specific quantitative corrections to thermality have been 
established~\cite{perthree,vijayherman}.  
More generally, it is far from obvious that many approximately
thermal states, with the same thermodynamic properties, cannot correspond
to many different microscopically defined
states differing in high-order,
hard-to-calculate correlations: generically, of course, they do!

More profound, perhaps, are more detailed dynamical arguments
tending toward the same conceptual tension.
Without
attempting to do these full justice, we can
paraphrase their common core as follows~\cite{sv21,thooftone,stu,polch}.  
Since black holes are intrinsically structureless, or
nearly so, well outside the horizon -- ``black holes have
no hair'' -- the process of collapse followed by slow
evaporation cannot be described by reversible, deterministic
equations.  For at intermediate stages -- after the collapse and
during the slow evaporation -- the structureless black hole
forms an extremely narrow information channel,
which cannot accommodate a complete, accurate record
of the initial state.   For purposes of this argument it
seems to be enough that there be no significant structure well
outside the horizon, because that space-time region predominantly
determines the asymptotic final state.
Banks and O'Loughlin~\cite{banks} especially
forcefully and explicitly advocated
that a distinction must be made between states, measurable
in the low energy theory, that should be counted towards the 
Bekenstein--Hawking entropy and other states, hidden from 
the observer strictly outside the hole, that should not.

Recently there have been some remarkable developments in
the microscopic (quantum) theory of black holes in string theory.
Early suggestions to identify certain electrically charged fundamental
string states with extremal black holes, with vanishing
classical area~\cite{duff}, led to a semi-quantitative matching of
the number of microscopic states with the area of an appropriate
stretched horizon~\cite{speculations,sen95,peet} and their remarkably
concrete realization~\cite{callan,waldram}.  
Subsequent suggestions and arguments~\cite{structure} that a
corresponding study of dyon states corresponding to extremal black
holes with non-vanishing area~\cite{dyon,cfthair1,cygeneral} would 
enable a successful quantitative
comparison with the classic Bekenstein-Hawking formula
were confirmed~\cite{strom96a}
in a surprising and brilliant manner as a by-product of the 
identification of
D-branes with string solitons~\cite{polch95a} and development
of an impressive associated calculational
technology~\cite{polch96a,intersect,vafa95a,vafa95b,sen95b}.
The understanding was immediately extended to 
near-extremal~\cite{callan96a,strom96b} and general black 
holes~\cite{strom96d,branedemo} although these results remain more
controversial than the extremal case. Work along related lines
has uncovered a beautiful, but still quite mysterious, relation between
the quantization of 5--branes and the second quantization of 
strings~\cite{dvv96b,dvv96c}.

While these developments seem to us fundamentally
to alter, perhaps permanently, the intellectual
framework
for the theoretical discussion, their limitations
should not be underestimated.
They apply at present only to
very special, particular black holes in highly
idealized models of physics.
We shall discuss this sort of limitation, which may or may not be
a purely technical problem, in more detail later.  For present purposes,
it is most important to notice that these developments
primarily touch only the
set of issues we discussed in paragraphs 2-4 above,
and characterized as thermodynamic.
The specifically
D-brane methods of calculation are not couched in
ordinary space-time terms,
and do not immediately explain {\it why\/}
a simple, universal result -- namely, that of Bekenstein and Hawking --
for the number of states should emerge.
The dynamical issues, which may be the more profound ones, appear
to require additional, or different, considerations.

The present work takes steps
toward a microscopic evaluation
of black hole entropy which emphasizes, rather than bypasses,
space-time issues.
Our working hypothesis, motivated by a desire to avoid
the dynamical arguments noted above, is that the internal structure
of black holes,
and therefore ultimately their entropy
the thermal properties, must be accurately represented by classical hair.
We recognize that this is a radical hypothesis, and that it
is not true in a generic
effective field theory of gravity.  After all, there are some
powerful no-hair
theorems which apply to simple, but widely used,
effective field theories (e.g., the 
Standard Model)~\cite{nohair,carter70,wald}.
However, of course, the generic effective field theory of gravity
almost certainly does not correspond to the reduction
of an adequate quantum theory,
as indicated by the presence of ultraviolet
divergences.
String theory is an increasingly popular candidate to provide an
adequate quantum theory of gravity; and the effective field theories
derived from it are much larger and more elaborate than have been
commonly considered previously.
The problem of insuring deterministic evolution may provide
another,
quite different indication that a substantial
expansion of the dynamical arena is
in order.
Indeed, the loophole in the no-hair theorem we seek to
exploit is
intrinsically connected to the existence of additional dimensions.
Our classical hair is related to the vast degeneracy among
black holes that arises
because the macroscopic charges of the black hole, which specify it
uniquely to an observer who does not probe the compactified dimensions,
do not uniquely specify
the details of the configuration in the compactified dimensions.
Although concrete identification of the complete catalogue
of hair may be difficult in any particular case,
this sort of
construction appears to us sufficiently robust that it is a viable
candidate for black hole entropy in general.

Conventional wisdom argues, in
Kaluza-Klein reduction, that
excitations with non--trivial dependence on small internal 
dimensions are very massive, so that
only the constant mode is important
in the low energy field theory.
Although it invokes structure in the extra
dimensions in a crucial way, our proposal does not
contradict this conventional wisdom as it is usually applied.
Local probes can only detect the hair
if they are sensitive to the internal dimensions; since
such probes must inevitably involve heavy degrees of freedom,
there is a definite sense in which
the hair is decoupled from the low energy world.
However, if we are correct, these degrees of freedom do not
decouple from the information flow in black hole processes, and
must be kept if one is to obtain a closed (unitary) description.

It is a special virtue of this approach that the classical
hair is measurable, in principle, at asymptotic distances. 
This is quite relevant for the information problem because hair 
stretching to infinity is retained permanently in boundary conditions,  
and is therefore in no danger of being `lost' at intermediate stages.
Thus the existence of abundant classical hair transforms this problem
profoundly: retention of the information becomes the default option,
which could only be endangered by unexpected strong coupling effects.

\bigskip

In the companion to this paper a general, rather abstract formalism
embodying our proposal was developed ~\cite{hair1}.
In the present paper
we analyze a specific example, Cveti\v{c}--Youm dyon.
The paper is organized
as follows.  In the first section we write the Cveti\v{c}--Youm dyon
in its four-dimensional and ten dimensional forms and discuss the 
range of validity of these classical solutions. 
Next we turn to the explicit construction of classical hair. 
This was already outlined in the Appendix of~\cite{hair1} 
but here the solution is recast in canonical form and the effective 
surface theory identified explicitly.
The semiclassical quantization of the black hole then follows. 
As a result we find that both the tension and the constraints
on the effective string parametrizing the collective excitations 
of the black hole agree exactly with the analogous ones for a 
fundamental string.
We then discuss discuss global issues that arise in
counting classical hair states.  Finally we discuss the
relation of our program to other work on the subject and 
speculate briefly on the possibility of a more general and profound 
relation between classical hair and black hole entropy.
The appendix is devoted to the quantization of the macroscopic 
black hole parameters.

\section{The Cveti\v{c}--Youm dyon}

\label{sec:macro}
Our working example is the Cveti\v{c}--Youm~\cite{dyon}. 
This is a black hole solution of the equations of motion that 
result from toroidal compactification of $N=1$ supergravity in $10$ 
dimensions. This theory in turn is the low-energy field theory 
limit string theory. Various properties of the solution were discussed 
in~\cite{cfthair1,structure}. In its full glory it is
\bea
ds^2 &=&  -\lambda dt^2  + \lambda^{-1}dr^2 +
R^2 (d\theta^2 + {\rm sin}^2\theta d\phi^2) \nonumber \\
\lambda &=& [ (1+{ {\bf Q^{(1)}} \over r}) (1+{ {\bf Q^{(2)}} \over r})
(1+{ {\bf P^{(1)}} \over r}) (1+{ {\bf P^{(2)}} \over r}) ]^{-{1\over 2}} 
\nonumber \\
R^2 &=& r^2 [ (1+{ {\bf Q^{(1)}} \over r}) (1+{ {\bf Q^{(2)}} \over r})
(1+{ {\bf P^{(1)}} \over r}) (1+{ {\bf P^{(2)}} \over r}) ]^{1\over 2} 
\nonumber \\
e^{2\phi}&=&\left[ {( 1+{ {\bf P^{(1)}} \over r})(1+{ {\bf P^{(2)}} \over r })
\over (1+{ {\bf Q^{(1)}} \over r })(1+{ {\bf Q^{(2)}} \over r }}
 \right]^{1\over 2} e^{2\phi_\infty} \nonumber \\
G_{44}&=&\left( { 1+{ {\bf P^{(2)}} \over r} \over
1+{ {\bf P^{(1)}} \over r }} \right)G_{44\infty}~~;~~~
G_{99}=\left( { 1+{ {\bf Q^{(1)}} \over r} \over
1+{ {\bf Q^{(2)}} \over r }} \right)G_{99\infty} \nonumber \\
A_\phi^{(1)}&=&P^{(1)} (\pm 1-\cos\theta )~~
;~~~A_\phi^{(7)}= P^{(7)}(\pm 1-\cos\theta ) \nonumber \\
A_t^{(6)}&=& -{Q^{(6)}\over r+{\bf Q}^{(1)}}~~
;~~~A_t^{(12)}=-{Q^{(12)}\over r+{\bf Q}^{(2)}} 
\label{eqn:solution}
\eea
The black hole is parametrized by four independent charges, two
electric and two magnetic. Their screened (boldfaced) incarnations 
have dimension of length, and are related to the physical charges 
through
\be
({\bf P^{(1)}},{\bf P^{(2)}},{\bf Q^{(1)}},{\bf Q^{(2)}})=
{\sqrt{\alpha^\prime} } (G_{44\infty}^{1\over 2}P_1,
G_{44\infty}^{-{1\over 2}}P_7,G_{99\infty}^{1\over 2}Q_6,
G_{99\infty}^{-{1\over 2}}Q_{12} )
\ee
In the remainder of the paper we take $G_{44\infty}=G_{99\infty}=1$
for simplicity in notation. The appendix of this paper is devoted 
to a discussion of the normalization and quantization of the $U(1)$ charges.

The black hole is the 4 dimensional form of a solution to the 
underlying 10 dimensional theory. Using this correspondence 
(given precisely in the appendix eq.~\ref{eqn:kaluzaklein}) 
the line element can be written in the obscure, but nevertheless 
convenient form
\bea
dS^2 &=& F du ( dv + K du ) +G_{ij}dx^i dx^j \nonumber \\
G_{ij}dx^i dx^j  &=& f [ k 
( dx^{(4)} + {\bf P^{(1)}}(1-\cos\theta)d\phi)^2
+ k^{-1}( dr^2+ r^2 (d\theta^2 +\sin\theta^2 d\phi^2 ))] \nonumber \\
&+& \sum_{i,j=5}^{8} \delta_{ij}dx^i dx^j 
\label{eqn:stringmetric}
\eea
in string metric. Here $u=x^{(9)}-t$, $v=x^{(9)}+t$, and
\be
F^{-1}=1+ {{\bf Q^{(2)}}\over r}~~;~~~K= {{\bf Q^{(1)}}\over r}~~;~~~
f=1+ {{\bf P^{(2)}}\over r}~~;~~~k^{-1}=1+{ {\bf P^{(1)}}\over r}
\label{eqn:fkfk}
\ee
The other nonvanishing fields are
\be
B_{uv}=F~~;~~~B_{\phi 4}={\bf P^{(2)}}(1-\cos\theta)~~;~~~e^{\Phi}=Ff
\label{eqn:bbdil}
\ee
This is the form of the black hole that appeared 
in eq. I.A.1\footnote{The roman numeral I refers to equation 
numbers in~\cite{hair1}. The A is a reference to the Appendix of 
that paper.}. It is repeated here for ease of reference.

We will work in the limit where string theory reduces to a 10 dimensional 
field theory. This is only reasonable if all scales are much larger than the 
string length $\sqrt{\alpha^\prime}$. Specifically, it would be dubious 
to locate the horizon more precisely than the string length. However, physical 
distances $\rho$ are related to coordinate distances by 
$\rho\simeq {\bf P}^{(1)} {\bf P}^{(2)}{\rm ln}~r$ close to the horizon 
at $r=0$; so this requirement concerns exponentially small coordinate 
distances and will not be relevant in this paper. It should
be emphasized that this argument relies on the kinematics of 
extremal black holes only; so it is valid for a much wider class of black 
holes than the one considered here. 

As explained in the introduction, black holes suffer from well-known 
conceptual problems that rely on one's
having great confidence in the classical 
metric, even for their formulation. 
The persistence of trustworthy geometric data means that these
problems still must be addressed.

\section{The Classical Hair and the Constraints}
\label{chap:explicit}

Classical hair on the Cveti\v{c}--Youm dyon was exhibited already
in the appendix of~\cite{hair1}. 
The purpose of this section, and the following, is to analyze classical
hair using the canonical formalism developed in~\cite{hair1}.
This provides a concrete example that complements the
abstract formulae. 

For the black hole without hair we choose the Schwarzchild time $t$ 
as canonical time and split eqs.~\ref{eqn:stringmetric}--~\ref{eqn:bbdil} 
in spatial and temporal parts, following~\cite{hair1}. 
The fields become
\bea
dS^2 & =& g_{99}~ (dx^9)^2 +g_{ij} dx^i dx^j \nonumber \\
g_{99} & =& F(K+1) \nonumber \\
N_9 &=& -FK \nonumber \\
N & = &\sqrt{F\over K+1} \nonumber \\ 
B_{t9}&=& -F 
\label{eqn:gaugefcts}
\eea
Here the transverse spatial metric $g_{ij}$ is identical to the
transverse spacetime metric $G_{ij}$ in eq.~\ref{eqn:stringmetric}.
The additional non--zero fields ( $B_{\phi 4}$ and $e^\Phi$ )
will not be needed.

The canonical momenta can be calculated from eq. I.2.5 . 
The non--vanishing components are
\bea
{\cal E}^{r9} &=& {1\over 2}{\bf Q}^{(2)} \nonumber \\
\Pi^r_{9} &= &{1\over 2}r^2 \partial_r K 
\label{eqn:expmomnm}
\eea
It is straightforward (but tedious) to verify explicitly 
that the fields in eqs.~\ref{eqn:gaugefcts}--\ref{eqn:expmomnm}~solve 
the constraints eqs. I.2.8 -- I.2.10 , 
as they should. 

We are interested in a solution that is modified due to
the presence of classical hair, and want to isolate
the dynamics of the hair, following the strategy of sec. I.4 .
Classical hair coordinates of the form identified previously are
assumed, and their conjugate momenta follow from the constraints.
For each canonical pair of background fields we can specify the 
field freely, but we must allow its conjugate momenta to respond
to the presence of hair. Inspired by the backreaction in 
eq. I.A.11 , found using the Lagrange formalism,
we will find it convenient that, 
for the $g_{99}$ field, we specify instead $\Pi^{99}=0$
and allow the field to depend on the hair. (This just amounts to a
canonical redefinition of variables.) Now,
the constraints eqs. I.2.8 -- I.2.10
are rather complicated non--linear differential equations
and we must solve for the momentum (and $g_{99}$ ) .
The strategy will be to anticipate the momenta 
eq.~\ref{eqn:expmomnm} found in the absence
of hair and introduce non--trivial components 
only when the constraints force it.

\subsection{Gauge Hair}
We will only consider gauge hair. 
The analogous calculation for Goldstone hair is in progress. 
In considerations later in the chapter we will assume, for purposes
of discussion, 
that the result for gauge hair extends to other kinds of hair,
though this is not at all obvious and may not be true.

 From eq. I.A.9 we expect hair
of the form
\be
F_{rt}= -F_{r9} = \partial_r F ~q^{\prime} = 
{{\bf Q}^{(2)}\over ({\bf Q}^{(2)}+r)^2}~q^{\prime}
\ee
We suppress the internal index $I=1,\cdots,16$. 
In the canonical formalism the function $q^\prime$ is specified 
as a function of $x^9$ at a fixed time slice; so the time 
dependence will only appear when the equations of motion are
considered in sec.~\ref{sec:eom}. 
The prime denotes derivative with respect to $x^9$, and
we should note that $q$ is normalized differently than 
in~\cite{structure}. We choose the gauge $A_9=A_t=0$ and find
\be
A_r = \partial_r F~q
\label{eqn:argauge}
\ee
The momentum ${\cal E}^9$ is non--dynamical because it
is conjugate to $A_9=0$. It can
be verified by writing out eq. I.2.9 that it enters 
the Hamiltonian as
\be
H = {1\over 2}Ng^{-{1\over 2}} e^{2\Phi}g_{99}
({\cal E}^9 +{\cal E}^{9r}A_r)({\cal E}^9 +{\cal E}^{9r}A_r)
+ {\rm terms~independent~of~{\cal E}^9 }
\ee
We can impose the corresponding equation of motion as a
rigid (first class) constraint by making the identification
\be
{\cal E}^9= -{\cal E}^{9r}A_r
\ee
in the following. Noting that ${\cal E}^{r9}$ is a constant,
antisymmetric in the two indices,
the Gauss' law constraint eq. I.2.8 
\be
\partial_\alpha ({\cal E}^{\alpha}-{\cal E}^{\alpha\beta} A_\beta )
=\partial_9 ({\cal E}^9-{\cal E}^{9r} A_r )+
\partial_r ({\cal E}^r-{\cal E}^{r9} A_9 )=0
\ee
can be integrated to
\be
{\cal E}^r  =-2{\cal E}^{r9}F~q^\prime 
= -{ {\bf Q}^{(2)}\over (1+{ {\bf Q}^{(2)}\over r})}  q^\prime
\label{eqn:erexplicit}
\ee
A constant of integration was undetermined {\it a priori}.
With the choice eq.~\ref{eqn:erexplicit} the momentum agrees 
with the on--shell 
result that can be obtained from eq. I.2.5 .

Now consider the Hamiltonian constraint eq. I.2.9.
The gravitational momenta
and the metric function $K$ are unknowns. Expanding the
curvature symbol and writing out the electromagnetic terms
we find
\bea
{\cal H}&= &g^{-{1\over 2}}e^{2\Phi}
[{\rm Tr}~\Pi^2 +{1\over 2}\Pi^\Phi {\rm Tr}~\Pi +{D-2\over 16}\Pi^2_{\Phi}]
- - g^{1\over 2}e^{-2\Phi}g^{rr}{1\over 2(K+1)^2} (\partial_r K)^2 \nonumber \\
&+ & g^{1\over 2}e^{-2\Phi}g^{rr} {1\over K+1} 
[{1\over F} (A^\prime_r)^2 + {1\over r^2}\partial_r (r^2 \partial_r K) ] 
= 0 
\label{eqn:explham}
\eea
after a lengthy calculation.
Requiring the last parenthesis to vanish by itself
gives $K$ in terms of the hair 
\be
K = {{\bf Q}^{(1)}\over r} - {{\bf Q}^{(2)2}\over 2r (r+{\bf Q}^{(2)})}
(q^\prime )^2 
\label{eqn:explk}
\ee
The constant at infinity was determined by the
prescribed charge of the black hole. This expression agrees 
with eq. I.A.11 . The other terms in 
eq.~\ref{eqn:explham} will be considered below.

Next we consider the supermomentum constraints ${\cal H}_\alpha = 0$
arising from eq. I.2.10 . Recall that the gravitational 
momenta are tensor densities so that
\be
\Pi^{\alpha\beta}_{~~|\beta} = \Pi^{\alpha\beta}_{~~,\beta}
+ \Gamma^{\alpha}_{~\beta\gamma}\Pi^{\beta\gamma}
\ee
Assuming that only $\Pi^r_9 \neq 0$ the ${\cal H}^9 =0$ constraint
becomes
\be
\Pi^{9\beta}_{~~|\beta}=g^{99} \partial_r (g_{99}\Pi^{9r})
={1\over 2}F^{9\alpha} ( {\cal E}_\alpha +
{\cal E}_{\alpha\beta}A^\beta ) =
 {1\over 2}g^{99}\partial_9 A_r ~{\cal E}^r
\ee
and the momentum is found to be
\be
\Pi^r_9 = - {{\bf Q}^{(2)}\over 4(1+{{\bf Q}^{(2)}\over r})^2} (q^\prime)^2
+ {\rm constant}
= {1\over 2}r^2 \partial_r K
\label{eqn:pir}
\ee
The constant of integration is independent of $r$ but could depend
on $x^9$. It was determined so that the first line 
in eq.~\ref{eqn:explham} vanishes without forcing additional 
non--zero momenta . 
The formula for $\Pi^r_9$ eq.~\ref{eqn:pir}~is formally 
identical to eq.~\ref{eqn:expmomnm}, which is valid when there is no hair, 
but the expression eq.~\ref{eqn:explk}~for $K$ now depends on the hair.

With a certain exception, discussed in sec.~\ref{sec:final},
the remaining supermomentum constraints 
${\cal H}_\alpha=0$ are not affected by the gauge hair.


\subsection{The Final Constraint}
\label{sec:final}
If only the hair variables depended on $x^9$ we would at this
point have checked all the constraints. However the
backreaction on the geometry is taken into account by 
allowing $g_{99}$ to depend on the hair variables, 
and non--trivial dependence of the metric on $x^9$ ensues. By
inspecting the appropriate expansion of the
Ricci scalar one can show that the Hamiltonian constraint remain
satisfied if this dependence is allowed. However, since $\Pi^9_r$
in eq.~\ref{eqn:pir} depends implicitly on the hair, the supermomentum
constraint ${\cal H}^r=0$ require further consideration. 
The {\it ansatz}
\be
\Pi_{ij}= -{1\over 4}~g_{ij}~\Pi^\Phi ={1\over D-2}~g_{ij}{\rm Tr}~\Pi
\ee
can be interpreted geometrically as a non--zero extrinsic
curvature ${\rm Tr}~{\bf K}$, using eq. I.2.5 .
It is an easy but non--trivial calculation to verify that the 
Hamiltonian constraint remains satisfied.
Using the {\it ansatz} and the explicit form of the
metric we find 
\be
\Pi^{rj}_{~~|j}-{1\over 2}\Pi^\Phi \partial^r \Phi
={r^2\over F}\partial_r [{F\over r^2}{1\over D-2}
{\rm Tr}~\Pi ] = \Pi^{r9}_{~~|9}
\ee
with the last equality expressing ${\cal H}^r=0$. 
Using eq.~\ref{eqn:explk}
and the metric this can be integrated to
\be
{1\over D-2}{\rm Tr}~\Pi = - r^2 fk^{-1}{1\over 2F(K+1)}\partial_9 K
\ee
We choose the constant (that may depend on $x^9$ ) to vanish.

The appearance of $r^2fk^{-1}=(r+{\bf P^{(1)}})(r+{\bf P^{(2)}})$ is
a new and tantalizing feature. In all our previous calculations 
the magnetic charges canceled in final results. 
${\rm Tr}~\Pi$ vanishes on--shell, but
the off-shell value can be important for the quantization of the system.
In fact, as we shall see in sec.~\ref{sec:tension}, the tension 
renormalization anticipated in~\cite{structure} can be 
interpreted as failure of single 
valuedness in the $x^9$ variable. 
This phenomenon may be signaled by total derivative terms in the 
Hamiltonian density, 
and this is indeed how ${\rm Tr}~\Pi$ enters, as alluded 
to in the end of sec.~I.5 . A precise connection eludes us at this 
point, however, and in sec.~\ref{sec:tension} a different reasoning 
will be pursued.

\section{The Effective Surface Theory and Quantization}
\label{sec:efftheory}

In~\cite{hair1} properties of the effective theory
for the classical hair were derived in abstract form. Using the expressions 
from the previous section they can now be made explicit.

It was proposed in~\cite{hair1} that the condition 
\be
\Pi^r_\alpha =0~~~;~~~~\alpha\neq r
\ee
should be imposed at $r=0$. The only non--trivial condition arises
for $\alpha =9$ and it implements reparametrization invariance in the
9th direction. Using eq.~\ref{eqn:pir}~and eq.~\ref{eqn:explk}~we find
\be
{1\over 2}{\bf Q}^{(2)}~\langle ~( q^\prime )^2 ~\rangle  ={\bf Q}^{(1)}
\label{eqn:matchinga}
\ee
This relation has a profound consequence: it forces the existence of
hair. The physical interpretation is that the hair
must carry all the momentum ({\it i.e.}. charge, after compactification )
that is seen at infinity. The analogous condition for the fundamental
string was derived in~\cite{callan} by appealing to a cosmic censorship
hypothesis, and in~\cite{waldram} by insisting that the external
field matches on to a string source. It is interesting
that our requirement of no source agrees with the result that follows
from specific properties of a string source.

\subsection{Quantization of Gauge Hair}
In~\cite{callan,waldram} the condition eq.~\ref{eqn:matchinga}~was 
identified with the matching condition on the world sheet of
a string that acts as a source. Here we have no source and 
proceed instead to quantize the field directly. 
In the complete theory the Poisson brackets of the fields are
\be
\{ {\cal E}^r (\vec{x}) , A_r (\vec{x}^\prime) \} 
= 16\pi G_N ~{1\over 2}\delta (\vec{x}-\vec{x}^\prime )
\label{eqn:gaugepoisson}
\ee
(The normalization of the right hand side arises because
we have defined momenta as variations of $16\pi G_N L$.)
This induces brackets in the reduced theory.

The momentum conjugate to the hair variables is essentially
the $x^9$ derivative of the hair, as is characteristic of
chiral fields. The factor of ${1\over 2}$ on the right
hand side of eq.~\ref{eqn:gaugepoisson} is the notorious one
that arises in the quantization of chiral fields when the 
second class constraints are properly implemented.

Using the explicit expressions eq.~\ref{eqn:argauge}~ 
and eq.~\ref{eqn:erexplicit}~ we perform 
the radial integral and find
\be
4\pi\{ -{1\over 2}{\bf Q}^{(2)}q^\prime (x^{9}) , 
q (x^{\prime 9}) \} 
= 16\pi G_N ~{1\over 2}~L
\delta (x^9-x^{\prime 9} )
\ee
The trivial angular integral gave rise to the $4\pi$ on the left
hand side. The explicit volume factor $L=2\pi\sqrt{\alpha^\prime}$
was implied in the normalization of the $\delta$--function 
in $x^9$ and is necessary for $G_N$ to denote
the four--dimensional Newton's constant throughout.

We introduce Fourier modes 
\be
q(x^9)=\sum_n ~q_n~e^{ {2\pi in\over L}x^9}
\label{eqn:fourierm}
\ee
and find 
\be
{ 1\over 4G_N  }\{ -{1\over 2}{\bf Q}^{(2)}
 {in\over \sqrt{\alpha^\prime}} q_n , 
q_{n^\prime} \} = {1\over 2}\delta_{n+n^\prime}
\ee
Replacing the brackets with commutators, and
using eq.~\ref{eqn:fourierm} again, we find
\be
{ \sqrt{\alpha^\prime}\over 4G_N }
 ~{1\over 2}{\bf Q}^{(2)}~\langle ~( q^\prime )^2 ~\rangle
=N
\ee
Here the level is $N=\sum_n nN_n$ in terms
of occupation numbers of individual states $N_n$.  Now the matching condition 
eq.~\ref{eqn:matchinga}~ can be written
\be
N = { \sqrt{\alpha^\prime}\over 4G_N }{\bf Q}^{(1)}=
{2Q^{(1)}\over g^2_{\rm st} } = q^{(1)}
\label{eqn:matchingb}
\ee
in terms of the quantized charges of the Appendix. Both sides of this relation 
take all integer values.
It should be noted that, when expressed in terms of integers, 
the matching condition is independent of the proportionality
constant relating $G_N$ and $\alpha^\prime g^2_{\rm st}$  
despite the ubiquity of the parameters at intermediate stages.

The normalization in eq.~\ref{eqn:matchingb} agrees with the 
standard one obtained from world sheet considerations. 
Its physical origin is the requirement that the oscillators 
carry the total momentum is also related. 
The nonrenormalization of string tension discovered by Dabholkar 
and Harvey equates the tension of the fundamental string with the
tension that is measurable at infinity~\cite{dab89}. Here we go
a step further and treat the oscillations as {\it bona fide}
collective excitations which are quantized without any reference to
the microscopic theory.

\subsection{Equations of motion}
\label{sec:eom}
The reduced Hamiltonian eq. I.5.2  has the simple form
\be
16\pi G_N H_{\rm red} = 
2 \int_{r=0} \Pi^r_9 = - \int_{r=0} [{\bf Q}^{(1)}- {1\over 2}
{\bf Q}^{(2)}(q^\prime)^2 ]
\ee
The equations of motion are found by variation of the Hamiltonian. The 
relation between the variables $q^\prime$ that appear in this expression 
and the canonical variables appropriate for the variational principle 
were worked out in the previous sections. 
We find the simple expression
\be
 H_{\rm red} = - \int dx^9~~q^\prime~\pi_q
\ee
where we denoted by $\pi_q$ the canonically conjugate to $q$
and omitted a constant. The equations of motion become
\be
{\delta H_{\rm red}\over \delta\pi_q} = -q^\prime = \dot{q} 
~~;~~~~
{\delta H_{\rm red}\over \delta q} =  \pi^\prime_q = -\dot{\pi_q} 
\ee
where the dots denote time derivatives. The equations are easily
solved and a time dependence is found that demands
$q$, as well as $\pi_q$, to be functions of $x^9-t$.
We conclude that only left movers are solutions of the
equations of motion, as anticipated. 

The strategy is more important than the result: the 
hair variables were isolated from the background, in a canonical 
and off-shell fashion, before the equations of motion were found and
solved. It was presumably special properties of the example here 
that allowed simple, explicit expressions to be found;
but the calculation serves as a useful paradigm for the more
general case, discussed abstractly in~\cite{hair1}.

\subsection{No On-shell Area}
In sec. I.4 we gave a simple and general derivation of the black 
hole entropy from the macroscopic view point. It was 
the second term in the reduced Hamiltonian eq. I.2.10
the was responsible for entropy in the general setting. In the
current explicit example it has the form
\be
H_{\rm red} = - {\pi\over G_N}\Theta \sqrt{  {\bf P}^{(1)}
{\bf P}^{(2)}{\bf Q}^{(2)}[{\bf Q}^{(1)}- {1\over 2}
{\bf Q}^{(2)}(q^\prime)^2 ]}
\ee
It was omitted above because the temperature $\Theta=0$ for the 
extremal black hole under consideration. However, the form
of this term makes manifest some properties of the formalism
that are presumably generic. First of all it emphasizes 
the feature that the area is dynamical; so within the reduced
Hamiltonian formalism this term, and specifically the temperature, 
is expected to enter the equations of motion explicitly.

It is intriguing that the term in the reduced Hamiltonian 
which, in the absence of hair, quite explicitly can be interpreted 
as the entropy is dramatically modified in the presence of hair. 
Indeed, if we impose the matching condition eq.~\ref{eqn:matchinga}, 
the dynamical area of the black hole vanishes! This is gratifying,
because if the area were non--zero the black 
hole with hair -- supposedly a microscopically specified
state -- would itself be a thermodynamic body with internal 
structure~\footnote{This was pointed out to us
by A. Sen.}.  This motivation  appears so compelling, that after having
observed the phenomenon in a specific example, we are led to expect
a vanishing area for the general microstate of any black hole.
We should emphasize in this connection
that it is the modulus $g_{99}$ that shrinks 
close to the horizon, with the spatial geometry in the uncompactified 
dimensions exhibiting no conspicuous features. 

It is interesting to speculate on possible
dynamical consequences~\footnote{This
paragraph was the result of a discussion with S. Mathur.}. 
Consider an infalling observer who may or may not be
translationally invariant in the $x^9$ direction. In a given
microstate the black hole has structure in the $x^9$ direction, 
so the impinging observer encounters a potential, because the
equations are non-linear, and will generically 
develop structure too.  Of course 
whenever an incoming wave experiences a non--trivial potential, there is
a reflected wave, and as the modulus $g_{99}$ shrinks
to zero the relevant effective potential appears to become 
arbitrarily strong, and reflection complete.  
This suggests a mechanism ---
realizable within the semiclassical approximation --- 
that avoids information loss in a very real sense: it becomes 
impossible 
to fall into the black hole. This phenomenon,  an infinitely high
effective barrier that reflects incoming waves, was 
previously encountered 
for a certain special class of black holes by Holzhey and 
Wilczek~\cite{holzhey}. 

\section{Tension Renormalization}
\label{sec:tension}
A satisfactory understanding of the internal structure of black holes 
must include a quantitative agreement between the number of microstates
and the entropy that follows from macroscopic considerations. Using 
the statistical mechanics of ideal gasses in two dimensions, the matching 
condition eq.\ref{eqn:matchingb}, can be translated into an entropy
\be
S \simeq 4\pi\sqrt{ q^{(1)}}
\label{eqn:unrenent}
\ee
This microscopic result is much smaller than the macroscopic one
from the appendix eq. \ref{eqn:bhent}, which contains the product
of all four charges (two electric, two magnetic). The discrepancy lead
Horowitz and Marolf to conclude that the classical hair form mesoscopic 
structure, unrelated to the true microstructure that account for 
the black hole entropy~\cite{marolf1,marolf2}. Our interpretation 
is quite the contrary: the qualitative picture is viable but the
counting is incomplete because global issues have been slighted.
 
\subsection{Winding Strings}
For a first indication of this possibility note that the
matching condition eq.~\ref{eqn:matchingb} differs
from the standard perturbative matching condition
by a factor of $q^{(2)}$. This discrepancy derives directly 
from the canonical brackets eq.~\ref{eqn:gaugepoisson}. 
The arguments of the fields and the $\delta$--function on the right hand side
include $x^9$, but in string theory the spacetime coordinate
is related to the world sheet coordinate $\sigma$ by a multiplicative
integer winding number $w$, {\it i.e.} $x^9=w\sigma$. 
The winding number couples to the Kalb--Ramond field, characterized
by the quantum $q^{(2)}$; so we are led to identify $w$ 
with $q^{(2)}$ and use the variable $\sigma={1\over q^{(2)}}x^9$ 
in the canonical brackets and the Fourier transform. Then the 
perturbative matching condition discussed above, {\it viz}.
\be
N = q^{(1)}q^{(2)}
\ee
is then neatly recovered.

Considering the physical nature of a string it is perhaps unsurprising
that there is a winding number. However, an observer of the 
field created by a string will detect momenta that are smaller 
than is normally allowed in a volume with the dimension
of the compactified space. These small momenta are quite
puzzling, but certainly real and measurable, even at infinity.
Their physical manifestation is the failure of recovering measured values
of the fields upon circumnavigating the compactified dimension.
To parametrize this, the observer may adopt a new definition of 
the dimension of the compactified space ( $L\rightarrow q^{(2)}L$ ) or, 
alternatively, conclude that in this environment 
the string tension is larger than its universal value 
$T={1\over 2\pi\alpha^\prime}$. The phenomenon of small momenta
is referred to as tension 
renormalization~\cite{speculations,structure,mathur96,susskind96}.
The concept has recently been made mathematically precise~\cite{dvv96d} 
with a relation to the winding sectors of the orbifold appearing 
in~\cite{strom96d}.

As an aside, consider a component of the field with momentum 
that is larger than the minimal
value (by a factor that is not a prime). Contributions to the 
amplitude arise from individual, maximally wound strings with 
this momentum; but also from multiple string configurations 
that each wind less. In the underlying Hilbert space of string 
field theory a distinction should presumably be made between 
such contributions, and this may affect the proper counting 
of degeneracy. However, it would seem very difficult for the observer
far away to discern such subtle differences, even though
careful experiments using Aharonov--Bohm type correlations might
conceivably accomplish it. Moreover, since the configurations
in question have identical classical actions it is possible that, 
even in principle, they should not be distinguished when 
calculating the volume of classical phase space that should be compared 
to the Bekenstein--Hawking entropy. Fortunately the leading 
order result for the entropy does not seem to depend on this issue, 
and we can simply assume that a single effective string suffices.

\subsection{Magnetic Flux}

At this point we have recovered the well-known degeneracy of
the fundamental string from a spacetime point of view.
In other words we have found a non--renormalization result 
for the tension that expresses the absence of gravitational backreaction
on the classical hair. It seems, alas, that the classical hair 
considered here is simply the external field of a fundamental 
string propagating in an inert black hole background, rather than an 
effective string theory that parametrize collective excitations 
of the black hole. However, as it currently stands the 
counting gives a result that is inconsistent with duality,
a symmetry of the low energy theory.  This symmetry interchanges
electric and magnetic charges, but leaves the degeneracy of states
({\it i.e}. here the entropy) invariant.  So we know the 
understanding must be incomplete. 

First reconsider how the winding emerges technically:
it amounts to the appearance of a phase in the 
hair variables when they are taken around the $x^9$ direction. 
Then it can be concluded that the Fourier transform 
eq.~\ref{eqn:fourierm} uses a length $L$ which is too small, 
because 
only when the effective length 
is increased do proper single valued fields arise.
 
The possibility of this kind of phase can be understood
by reasoning along the lines of sec.~\ref{sec:macro} , as follows. 
In a background of charge $q^{(2)}$, gauge transformations 
with gauge function $\Lambda$ of minimal period $2\pi$ change the
action by $2\pi q^{(2)}$. A transformation that formally
acts like a gauge transformation with gauge function $\Lambda$ 
of period $2\pi/q^{(2)}$ generates new solutions 
that are {\em not} gauge equivalent to the original one. 
The field theory of classical hair allows Wilson lines created 
by this kind of transformation to be attached to each hair 
independently. The mechanism proposed here to generate 
more classical hair, once we have discovered some, 
is closely related to the construction of discrete gauge 
hair in~\cite{qhair}. All winding modes can be accounted for 
this way.
 
We now turn our attention to the role of the magnetic charges. 
Again the background allows `small' gauge transformations 
that generate new solutions from existing ones. 
 From the requirement that the Dirac--string remains hidden
the allowed periodicities are ${2\pi\over p^{(1)}}$ 
( or ${2\pi\over p^{(2)}}$ for the other gauge field ).
Again we can apply this kind of transformation on each of 
the modes of the classical hair independently. In consequence,
configurations exist that, as we move around the 9th direction,
change by a phase. If $p^{(1)}$, $p^{(2)}$, and $q^{(2)}$ are
relatively prime the effective length, needed to obtain a
single-valued field, becomes the product $q^{(2)}~p^{(1)}~p^{(2)}$. 
With this tension renormalization the correct matching condition 
entropy are recovered.

To put this heuristic argument on a firmer basis it must be verified 
more explicitly that the appropriate couplings are present, for
all modes of the classical hair, and a better understanding of
the field theoretic basis for independent transformations on
each mode is also needed.  Though
much remains unsettled, we do think it is now quite plausible
that the right sort of field
theory contains the features necessary for a quantitative 
account of the black hole entropy using classical hair, once global
issues are properly addressed.


\subsection{Relation to Other Work}

Cveti\v{c} and Tseytlin~\cite{cfthair1} (see
also~\cite{cfthair2,newtseytlin} )
noted that the classical hair can be represented as marginal
deformations in the underlying conformal field theory. In this
approach the appropriate tension renormalization appears naturally 
with the predicted coefficient, as the Kac--Moody level of 
the current algebra.

The left hand side of the matching condition is the
level of $c=24$ physical oscillators of heterotic string theory.
As explained in the Appendix the formula for the
black hole entropy depends on the spectrum of
allowed magnetic charges. If Dirac quantization and
a matching condition with minimal charges are used, as seems most 
reasonable, there is a discrepancy of a factor of 2 between the
renormalized version of eq.~\ref{eqn:unrenent} and the
Bekenstein-Hawking result eq.~\ref{eqn:bhent}. Instead
$c=6$ is required to bring the renormalized the results into 
agreement.


Tseytlin noted that $c=6$ could arise naturally from classical hair 
of the type presented here if only the Goldstone hair in the 
$(r,\theta,\phi,4)$ directions suffer tension 
renormalization~\cite{newtseytlin}.  (The contribution of fermionic 
partners presumably follows automatically if spacetime supersymmetry 
is realized, although we have not considered this carefully within our
framework.)
Unfortunately the modes in the $(r,\theta,\phi,4)$ are precisely the
ones we have not been able to handle analytically.

Strominger and Vafa presented a calculation in the type II 
theory that represents the microscopic states of certain
5 dimensional black holes as an effective string that
indeed has $c=6$~\cite{strom96a}.
The effective string that appears in this accounting for black hole 
entropy is not a renormalized version of the fundamental string, but rather 
some new kind of chiral string that has $c=6$, and is confined to 
live on 6--dimensional world volumes. It was proposed by Dijkgraaf, 
Verlinde, and Verlinde that this string should be 
taken as a serious starting point for quantization~\cite{dvv96a,dvv96b}. 
 
Translating this reasoning into our spacetime picture
we are lead to expect that, in addition to the classical hair 
we have exhibited explicitly, there are some excitations that
depend on the $(x^5,x^6,x^7,x^8)$ variables as well as $x^9-t$ and 
that only those species suffer tension renormalization. 
Such classical 
solutions would be consistent with supersymmetry and therefore quite
plausibly exist, but their explicit construction may be difficult.


\section{Discussion}
\label{sec:waiver}
At the core of our program, to account for black hole entropy using
classical hair, is a connection between global constraints a
field configuration -- here, that it contains a black hole with
given macroscopic quantum numbers -- and some characterization
of the number of solutions which
realize it.  This is a theme that has occurred
before in physics and mathematics, as the connection between anomalies
and zero-modes, or as index theorems~\cite{jackiw}.   
Here we seek an index theorem with a new element, in that the number 
of solutions is quantified as a volume -- the volume of
classical phase-space.  If such an index theorem were found, it would
for many purposes free us from the necessity of actually finding the
explicit solutions, which as we have seen is a painful process at best.
Ideally, it would relate an `anomaly' in information flow across the
horizon to the build-up of classical hair outside, and thus could be
interpreted as a necessary
physical consistency condition on the effective theory.  At present,
however, these ideas are no more than attractive speculation.

Our concept of classical hair that is measurable far away implies 
that all information is contained in scattering amplitudes that 
are susceptible only to the leading behavior of the potentials, {\it i.e.} 
small angle scattering. This appears closely related to the program 
pursued by Mandal and collaborators~\cite{mandal1,mandal2} and to 
the calculations of Das and Mathur~\cite{mathur96,mathur96b}.
On the other hand we see no obvious connection with
the statistical hair advocated by Strominger~\cite{strom96e} or the
quantum hair discussed by Banks~\cite{banks96}.

't Hooft has advocated for some time that the consistency of
black hole quantum mechanics
severely constrains the form of the underlying microscopic theory
~\cite{thooft}.
We hope to have made it
plausible that black hole entropy can be accounted
for microscopically, but only in special many-dimensional theories
with appropriate symmetries.
This seems to embody a part of 't Hooft's program. 


{\bf Acknowledgments}
FL would like to thank V. Balasubramanian, M. Cveti\v{c},
J. Maldacena, S. Mathur, A. Sen, and A. Tseytlin for discussions and
NORDITA for hospitality while this manuscript was completed.  FL would
also like to thank Princeton University where the bulk of this work
was carried out. We would also like to thank G. Mandal for a
stimulating discussion.


\newpage

\appendix

\section{Quantization of Macroscopic Charges}

This Appendix contains an elementary discussion, within the
framework of field theory, of the quantization conditions on 
the $U(1)$ charges. 

Consider dilaton gravity in ten dimensions, coupled to an antisymmetric
vector field. After toroidal compactification to 4 dimensions
the Lagrangian becomes

\bea
L &=& {1\over 16\pi G_N}\int d^4 x \sqrt{-g}
[R_g- 2\partial^\mu\phi\partial_\mu\phi \nonumber \\
&-&{\alpha^\prime\over 4}e^{-2(\phi-\phi_\infty )}
\sum^{6}_{a=1}(F_{\mu\nu}^{(a)}G^{aa}F^{(a)\mu\nu}
+F_{\mu\nu}^{(a+6)}G_{aa}F^{(a+6)\mu\nu})  \nonumber \\
&+& {1\over 4}\partial^\mu G^{ab} ~\partial_\mu G_{ab}  ] 
\label{eqn:action}
\eea

The fields $F_{\mu\nu}^{(a)}~(a=1,\cdots,6)$ are Kaluza Klein
fields, i.e. dimensionally reduced components of the metric field,
and the fields $F_{\mu\nu}^{(a+6)}~(a=1,\cdots,6)$ are winding
fields that arise from the antisymmetric vector field.
The compactification scale $\sqrt{\alpha^\prime}$ sets the relative
scale of the terms. We have ignored possible moduli from compactification
of $B_{IJ}$ and $A^{(i)}_I$; so this toy model contains only 
the most basic features
of string induced gravity. 
For the heterotic string the full bosonic Lagrangian is given 
in~\cite{maharana}.

In the toroidal  compactification that leads to the 
Lagrangian eq.~\ref{eqn:action}~the Kaluza--Klein fields 
are introduced by the decomposition
\be
dS^2 = G_{\mu\nu}dx^\mu dx^\nu
+G_{mn}(dx^m +A_\mu^{(m)}\sqrt{\alpha^\prime}~dx^\mu)
(dx^n +A_\nu^{(n)}\sqrt{\alpha^\prime}~dx^\nu)
\label{eqn:kaluzaklein}
\ee
of the string metric $G_{IJ}=e^{2(\phi-\phi_\infty)}g_{IJ}$. 
Therefore the gauge invariance $A_\mu\rightarrow A_\mu +\partial_\mu \Lambda$
(for each of the gauge fields) is induced by reparametrization invariance 
of the appropriate internal dimension. The gauge function is only 
defined up to periodicity because the internal coordinate is periodic. 
Having adopted the convention that the moduli at infinity are unity
the radii of the tori are simply $\sqrt{\alpha^\prime}$ 
in the current conventions and the appropriate
identification becomes $\Lambda\equiv \Lambda+2\pi$. 

Consider a single Kaluza--Klein field 
\be
L= {1\over 8\pi g_{\rm st}^2}\int  e^{-2(\phi-\phi_\infty)}
F_{\mu\nu}F^{\mu\nu} 
\label{eqn:emlag}
\ee
It is normalized as in eq.~\ref{eqn:action}
if we use $G_N = {1\over 8}\alpha^\prime g_{\rm st}^2$ 
as a convention that defines the string coupling.
The action has a subtle gauge dependence that is due 
to boundary terms at infinity. 
Indeed, for solutions to the equations of motion,
\be
\delta L=
{1\over 2\pi g_{\rm st}^2}\int_{r\rightarrow\infty} 
F_{\mu\nu} \delta A^{\nu} dS^\mu 
= {2Q\over g^2_{\rm st}}\int \delta A_t dt
\label{eqn:deltalag}
\ee
where the charge was defined by the asymptotic behavior
$F_{rt}\rightarrow {Q\over r^2}$ as $r\rightarrow\infty$.

As variation in eq.~\ref{eqn:deltalag} we consider the pure gauge 
transformation $\delta A_t=\partial_t \Lambda$. Recalling that the $\Lambda$
admits periodic identifications and imposing the quantization 
condition that the Lagrangian similarly is well defined up to 
multiples of $2\pi$, we find the quantization
condition
\be
{2Q\over g^2_{\rm st}} = {\rm integer}\equiv q
\label{eqn:qquant}
\ee
This holds for each Kaluza--Klein field separately. 

For gauge fields that derive from dimensional reduction
of the Kalb--Ramond field $B_{IJ}$ the situation is 
less straightforward. $T$--duality
indicates that such charges must have the same quantization rule, and
this will be adhered to in the following. For a heuristic 
understanding of how this emerges in a spacetime approach, 
implement $T$--duality by large coordinate transformations 
that interchange time with one of the internal coordinates. 
On the world sheet this symmetry is related to modular invariance; 
and in field theory it presumably follows analogously 
from general covariance and unitarity. The effective periodicity 
of the Euclidian time plays the role of temperature;
so it is intriguing that it appears in the quantization conditions. 

The Dirac quantization condition on the magnetic charges is
also a consequence of the  compactification: the magnetic 
gauge potentials are
$
A_\phi = P(\pm 1-\cos\theta)
$
on the north and south hemisphere, respectively . 
On the equator they are related by
$ A_\phi = A^\prime_\phi +\partial_\phi \Lambda $~;
so $\Lambda = 2P\phi$ must be an acceptable transition function. 
The implied periodicities from this and from the compactification
are
$\Lambda \equiv \Lambda +4\pi P \equiv \Lambda +2\pi n $, so
\be
2P={\rm integer} 
\label{eqn:pquant}
\ee
is the Dirac quantization condition for the magnetic charge. 

In field theory there is no fundamental requirement that all integer 
magnetic charges are realized in a given theory. However, 
in recent years evidence has been accumulating 
that string theory indeed saturates Dirac's quantization
condition. Perhaps the most convincing argument appears in
type I string theory where solitonic objects
have explicit realizations as D--branes; and $T$--duality
ensures that any number of D--branes must be realized. In this
context it can be verified by direct calculation~\cite{polch95a}
(or from topological considerations~\cite{ibrane}) that the D--branes
indeed carry the minimum Dirac quantum.
The corresponding result in heterotic string theory then
follows as a prediction of duality. 

The quantization on magnetic charges can be understood from duality 
in the field theory limit, as follows~\cite{sen94}.
Impressive evidence for non--abelian duality has accumulated 
for $N=4$ supersymmetric gauge field theory (for a review 
see~\cite{harvey2}). Normalizing the $SU(2)$ gauge field
part of the Lagrangian as
\be
L= {1\over 16\pi }\int~[
 {4\pi\over e^2}{\rm tr}~W_{\mu\nu}W^{\mu\nu}+{\theta\over 2\pi}
{\rm tr}~\tilde{W}_{\mu\nu}W^{\mu\nu}]
\label{eqn:susylag}
\ee
where
\be
W^{(a)}_{\mu\nu}= \partial_\mu A^{(a)}_\nu -\partial_\nu A^{(a)}_\mu
+\epsilon_{abc}A^{(b)}_\mu A^{(c)}_\nu
\ee
the non--abelian duality is implemented by the $SL(2,Z)$ parameter
\be
\tau = {\theta\over 2\pi}+{4\pi i\over e^2}
\ee
For $\theta=0$ the $Z_2$ subgroup
${4\pi\over e^2}\leftrightarrow {e^2\over 4\pi}$
is a special case of the larger duality group. The spectrum of the
theory includes dyons with electric charges quantized so that
$\frac{4\pi}{e^2}Q$ is an integer and magnetic charges quantized
as integers. The spectrum satisfies Dirac's quantization condition with 
the magnetic quanta equal to twice their minimal values. This apparent 
doubling arises because both the electric and the magnetic charges are 
in the adjoint representation of the gauge group, rather than in the 
fundamental.

To find the quantization condition on heterotic string theory 
note that it develops non--abelian gauge symmetry at special points 
in the moduli space. At such points the $U(1)$'s in eq.~\ref{eqn:action} 
can be considered subgroups of $SU(2)$ gauge groups and the 
quantization conditions inferred from the non-abelian field theory. 
The result is extended to the rest of moduli space, by continuity. 

Concretely, consider a Kaluza--Klein field $F^{(1)}$ and a winding 
field $F^{(2)}$ and form $F^{(\pm)}=F^{(1)}\pm F^{(2)}$. 
The $U(1)$ field $F^{(-)}$ is part of a non-abelian gauge field and
we identify it with $W^{(3)}$. 
Now, two copies of the Lagrangian eq.~\ref{eqn:emlag} become
\be
L= {1\over 16\pi g_{\rm st}^2}\int  e^{-2(\phi-\phi_\infty)}
~[F^{(-)2}+F^{(+)2}]
\label{eqn:emlagtwo}
\ee
Regarding $F^{(+)}$ a fixed background field
and identifying $g^2_{\rm st}$ with ${e^2\over 4\pi}$ 
this agrees with an abelian subgroup of the Lagrangian 
eq.~\ref{eqn:susylag} for constant dilaton. 
The electric (and magnetic) charges of $F^{(1,2)}$, 
${1\over g^2_{\rm st}}Q^{(1,2)}$ ( and $P^{(1,2)}$ )
are quantized in half integers by eq.~\ref{eqn:qquant} 
( and eq.~\ref{eqn:pquant} ) so, keeping the charges of $F^{(+)}$ fixed 
at an even integer, the allowed electric (and magnetic) charges 
of $F^{(-)}$ are all the integers, in accordance with the 
non--abelian field theory\footnote{An odd $F^{(+)}$ charge introduces a shift
in the allowed $F^{(+)}$ charges by one half but the spacing remains 
the same.}. The string theory duality (with the minimum Dirac
quantum) therefore maps correctly to the non--abelian field theory 
duality (with twice the minimum quantum). The precise form of the 
embedding plays a crucial role in making the factors of two agree.


As an application of the quantization rules extract the ADM mass 
from the metric in eq.~\ref{eqn:solution}  
and apply the quantization rules on each charge, to find
\be
M_{\rm ADM} = {\sqrt{\alpha^\prime}\over 4G_N}
[Q^{(1)}+Q^{(2)}+P^{(1)}+P^{(2)}] = {1\over \sqrt{8G_N}} 
[g_{\rm st} (q^{(1)}+q^{(2)})+{1\over g_{\rm st}}(p^{(1)}+p^{(2)})]
\ee
Note that $G_N$ entered through boundary conditions at infinity that
define the ADM mass. This formula has several applications.
The invariance under inversion of the coupling constant and 
simultaneous interchange of electric and magnetic quantum numbers
can be taken as evidence for duality. Moreover, for states with no 
magnetic charge it agrees with
the standard world sheet expression if the integers
$q^1$ and $q^2$ are identified with the integer 
world sheet momentum and winding respectively. In either case the
consistency checks are independent of the proportionality constant relating 
$G_N$ and $\alpha^\prime g^2_{\rm st}$, and therefore of the
quantization condition.

The quantity of primary concern to us is the thermodynamic entropy 
of the black hole. It is
\be
S = {A\over 4G_N}=
{1\over 4G_N} 4\pi\alpha^\prime \sqrt{Q^{(1)}Q^{(2)}P^{(1)}P^{(2)}}
 = 2\pi \sqrt{q^{(1)}q^{(2)}p^{(1)}p^{(2)}}
\label{eqn:bhent}
\ee
Again $G_N$ entered through boundary conditions at infinity.
As discussed in~\cite{structure,kallosh96b}, the definition 
of the string coupling dropped out of this expression as did the
moduli (which are set to unity in the conventions here). It therefore 
relies on the quantization condition on the magnetic charges only.

\end{document}